\documentclass[
showpacs,
floatfix,
aps,
prb,
amsmath,
twocolumn,
superscriptaddress, 
tightenlines
groupaddress,
]{revtex4}

\usepackage{graphicx}
\usepackage{dcolumn}
\usepackage{amsmath}
\usepackage{amsfonts}
\usepackage{bm}
\usepackage{epsfig}
\usepackage{times}

\newcommand{\pc}{{\cal P}_{\omega}}
\newcommand{\be}{\begin{equation}}
\newcommand{\ee}{\end{equation}}
\newcommand{\bea}{\begin{eqnarray}}
\newcommand{\eea}{\end{eqnarray}}

\def\Edc{\mathcal{E}_{\mathrm{dc}}}

\def\eac{\epsilon_{\mathrm{ac}}}
\def\edc{\epsilon_{\mathrm{dc}}}

\def\oc{\omega_{\mbox{\scriptsize {c}}}}

\def\rc{R_{\mbox{\scriptsize {c}}}}

\def\tpi{\tau_{\pi}}

\def\tq{\tau_{\mbox{\scriptsize {q}}}}

\def\ttr{\tau_{\mbox{\scriptsize {tr}}}}

\def\ttr{\tau_{\mbox{\scriptsize {tr}}}}

\newcommand{\req}[1]{Eq.\,(\ref{#1})}
\newcommand{\reqs}[2]{Eqs.\,(\ref{#1}),(\ref{#2})}
\newcommand{\rfig}[1]{Fig.\,\ref{#1}}

\newcommand{\oncite}[1]{Ref.\,\onlinecite{#1}}

\begin{document}

\title{
Hall field-induced resistance oscillations in tilted magnetic fields
}

\author{A.\,T. Hatke}
\affiliation{School of Physics and Astronomy, University of Minnesota, Minneapolis, Minnesota 55455, USA}

\author{M.\,A. Zudov}
\email[Corresponding author: ]{zudov@physics.umn.edu}
\affiliation{School of Physics and Astronomy, University of Minnesota, Minneapolis, Minnesota 55455, USA}

\author{L.\,N. Pfeiffer}
\affiliation{Princeton University, Department of Electrical Engineering, Princeton, NJ 08544, USA}

\author{K.\,W. West}
\affiliation{Princeton University, Department of Electrical Engineering, Princeton, NJ 08544, USA}

\received{November 10, 2010}

\begin{abstract}
We have studied the effect of an in-plane magnetic field on Hall field-induced resistance oscillations in high mobility two-dimensional electron systems.
We have found that the oscillation frequency depends only on the perpendicular component of the magnetic field but the oscillation amplitude decays exponentially with an in-plane component.
While these findings cannot be accounted for by existing theories of nonlinear transport, our analysis suggests that the decay can be explained by an in-plane magnetic field-induced modification of the quantum scattering rate. 
\end{abstract}
\maketitle

%\section{Introduction}
High mobility two-dimensional electron systems (2DESs) exhibit an array of fascinating transport phenomena occurring in very high Landau levels where the Shubnikov-de Haas oscillations (SdHOs) are not yet resolved. 
Among these are several classes of magnetoresistance oscillations, such as microwave-induced resistance oscillations (MIROs),\citep{miro:exp2,mani:2005,yang:2006,miro:th,dmitriev:2009bs} phonon induced resistance oscillations (PIROs),\citep{piro:1} and Hall field-induced resistance oscillations (HIROs),\citep{hiro:2,vavilov:2007}, as well as their combinations.\citep{comb:2,khodas:2008} 
In a very clean 2DES, MIROs and HIROs can lead to exotic zero-resistance\citep{zrs:exp2} and zero-differential resistance states,\citep{hatke:2010s} respectively, which can be explained in terms of instabilities and formation of current domains.\citep{zrs:th}

MIROs are observed in linear-response magnetoresistivity when a 2DES is irradiated by microwaves and can be understood in terms of microwave-induced transitions between the Landau levels which lead to oscillatory photoresistivity:\citep{dmitriev:2009bs}
\be
\frac {\delta \rho_\omega}{\rho_0} \simeq - \eta \pi\eac {\pc} \lambda^2 \sin 2\pi\eac.
\label{miro}
\ee
Here, $\rho_0$ is the resistivity at $B=0$, $\eac = \omega/\oc$, $\omega=2\pi f$ is the microwave frequency, $\oc=e B_\perp/m^*$ is the cyclotron frequency, $B_\perp$ is the magnetic field normal to the 2DES, $m^*$ is the effective mass, $\lambda = \exp(-\pi/\oc\tq)$ is the Dingle factor, $\tq$ is the quantum lifetime, $\pc$ is the dimensionless microwave power, and $\eta$ is the scattering parameter which depends on temperature and type of disorder in the 2DES.\citep{dmitriev:2009bs}

HIROs are observed in differential resistivity $r\equiv dV/dI$ when a direct current $I$ is passed through a 2DES and the perpendicular magnetic field, $B_\perp$, is varied.
HIROs originate from short-range impurity-mediated transitions between Landau levels tilted by the Hall electric field, $\Edc=\rho_{H}I/w$ ($\rho_H$ is the Hall resistivity, $w$ is the sample width).\citep{vavilov:2007,khodas:2008}
In this scenario, a characteristic scattering event involves an electron which is backscattered off of an impurity. 
The guiding center of such an electron is displaced by the cyclotron diameter, $2\rc$, and when $2\rc$  matches an integral multiple of the real-space Landau level separation, $\hbar\oc/e\Edc$, the probability of such events is enhanced.
This enhancement gives rise to a maximum in the differential resistivity occurring whenever $\edc \equiv e\Edc(2\rc)/\hbar\oc$ is equal to an integer.
At $2\pi\edc \gg 1$, HIROs are described by:\citep{vavilov:2007,khodas:2008}
\be
\frac {\delta r} {\rho_0} \simeq  \frac{16}{\pi} \frac{\ttr}{\tpi} \lambda^2\cos 2\pi\edc,
\label{hiro}
\ee
where $\ttr$ is the transport lifetime and $\tau_{\pi}$ is the time describing electron backscattering off of impurities.\citep{vavilov:2007,khodas:2008}

%MIRO in tilted field
The effect of an in-plane magnetic field, $B_\parallel$, on MIROs has been recently investigated in two independent experiments.\citep{mani:2005,yang:2006}
In \oncite{mani:2005} the magnetic field $B$ was tilted away from the normal to the 2DES by an angle $\theta$ and in \oncite{yang:2006} $B_\parallel$ was applied independently of $B_\perp$. 
While both experiments agreed that the MIRO frequency is governed by $B_\perp$, \oncite{yang:2006} found that MIROs are strongly suppressed under $B_\parallel \simeq 0.5$ T while in \oncite{mani:2005} MIROs were essentially unchanged up to $\theta \simeq 80^\circ$ ($B_\parallel \lesssim 1.2$ T).
This controversy and the very fact that the suppression of MIROs observed in \oncite{yang:2006} was left unexplained indicates that the role of $B_\parallel$ is not understood.
It is therefore interesting and timely to examine how other classes of resistance oscillations, {\em e.g.} HIROs, respond to $B_\parallel$.

In this Rapid Communication we report on the effect of an in-plane magnetic field on Hall field-induced resistance oscillations in a high mobility 2DES.
We employ a tilted-field setup and observe that, similar to MIROs,\citep{mani:2005,yang:2006} the HIRO frequency depends only on the perpendicular component of the magnetic field, $B_\perp$.
We further find that with increasing tilt angle $\theta$, HIROs are strongly suppressed by modest $B_\parallel \sim 1$ T.
The observed suppression is nonuniform and depends on the oscillation order; the lower order oscillations decay  faster than the higher order oscillations with increasing $\theta$.
While the suppression of the HIRO amplitude by $B_\parallel$ cannot be readily explained by existing theories, we discuss our findings in the context of \req{hiro} and show that the suppression can be understood in terms of a $B_\parallel$-induced reduction of the quantum lifetime.
However, identifying the origin of such modification remains the subject of future studies.

%\section{Experimental Details}
While similar results have been obtained from a variety of samples, the data presented here were obtained from a Hall bar ($w=100$ $\mu$m) cleaved from a GaAs/Al$_{0.24}$Ga$_{0.76}$As 300 \AA-wide quantum well grown by molecular beam epitaxy.
After a brief low temperature illumination with a red light emitting diode, this sample had electron density $n_e\simeq 3.6 \times 10^{11}$ cm$^{-2}$ and mobility $\mu \simeq 1.0 \times 10^7$ cm$^2$/Vs.
The experiment was performed in a $^3$He cryostat at $T\simeq 1.0$ K and $I=100$ $\mu$A using a standard low-frequency lock-in detection scheme.

%\section{Raw Data}
%%%%%%%%%%%%%%%%%%%%%%%%%%%%%%%%%%%%%%%%%%%%%%%%%
%fig 1
\begin{figure}[t]
%\resizebox{0.5\textwidth}{!}{
\includegraphics{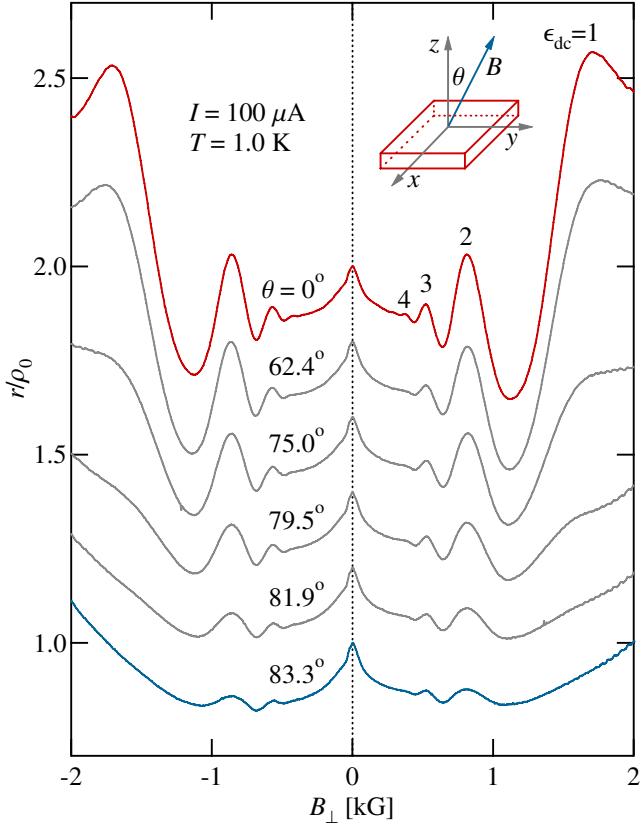}
%}
%\vspace{-0.1 in}
\caption{[Color online] Normalized differential resistivity $r/\rho_0$ vs $B_\perp$ for tilt angles $\theta \simeq 0^\circ, 62.4^\circ, 75.0^\circ, 79.5^\circ, 81.9^\circ, 83.3^\circ$.
The data are vertically offset for clarity in step of 0.2 starting from the highest $\theta$.
HIRO maxima are marked by $\edc=1,2,3,4$.
Inset: direction of $B$ with respect to the 2DES.
}
%\vspace{-0.15 in}
\label{fig1}
\end{figure}
%%%%%%%%%%%%%%%%%%%%%%%%%%%%%%%%%%%%%%%%%%%%%%%%%
In \rfig{fig1} we present the normalized differential resistivity $r/\rho_0$ vs $B_\perp$ for different $\theta$, as marked.
The data are vertically offset for clarity in step of 0.2 starting from the highest $\theta$.
At the top curve ($\theta=0$) the first four HIRO maxima are marked by $\edc=1,2,3,4$.
Examination of the data reveals that, regardless of $\theta$, the positions of the same-order maxima and minima roughly coincide with each other.
This finding provides firm experimental evidence that, similar to MIROs,\citep{mani:2005,yang:2006} the frequency of HIROs is controlled by $B_\perp$.
At the same time, it is evident that the oscillation amplitude decreases with increasing $\theta$. 

The top trace in \rfig{fig1} obtained at $B_\parallel = 0$ shows that the oscillation amplitude grows monotonically with increasing $B_\perp$  due to the increase of the Dingle factor $\lambda$ entering \req{hiro}.
Examination of the other data in \rfig{fig1} reveals that HIROs gradually decay with increasing $\theta$ and that the decay is nonuniform; the strongest, fundamental HIRO peak (cf.\,``$\edc=1$'') is much more sensitive to $\theta$ than the higher order peaks appearing at lower $B_\perp$.
Indeed, it virtually disappears at $\theta \simeq 81.9^\circ$ ($B_\parallel \simeq 1.1$ T) 
while the higher-order peaks are still clearly observed.
This finding indicates that at finite $\theta$ the amplitude is no longer a monotonic function of $B_\perp$ and thus cannot be described by an exponential dependence of \req{hiro}.

\begin{figure}[t]
%\resizebox{0.5\textwidth}{!}{
\includegraphics{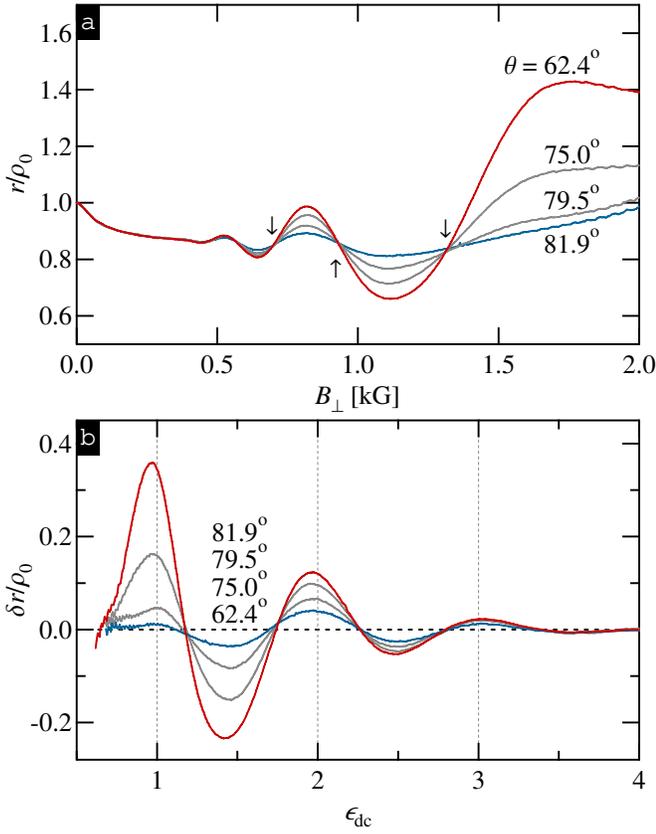}
%}
%\vspace{-0.15 in}
\caption{[Color online] (a) Normalized differential resistivity $r/\rho_0$ vs $B_\perp$ for $\theta =  62.4^\circ, 75.0^\circ, 79.5^\circ$, and $81.9^\circ$. The data are {\em not} vertically offset.
(b) Normalized oscillatory part of the differential resistivity $\delta r/\rho_0$ vs $\edc$ for the same tilt angles.
}
%\vspace{-0.15 in}
\label{fig2}
\end{figure}

Since \req{hiro} dictates that the HIRO amplitude $A \equiv (16/\pi)\cdot(\ttr/\tpi)\cdot\exp(-2\pi/\oc\tq)$, one has to examine possible effects of $B_\parallel$ on the various scattering parameters, {\em i.e.}, on $\ttr$, $\tpi$, and $\tq$. 
In our 2DES at weak $B_\perp$, $\rho \simeq \rho_0 \propto 1/\ttr$, and one can estimate the effect of $B_\parallel$ on $1/\ttr$ by investigating the evolution of the background part of $r$ with increasing $\theta$. 
This is done in \rfig{fig2}\,(a) showing $r/\rho_0$ as a function of $B_\perp$ for different $\theta$ (as marked) {\em without} a vertical offset.
Plotted in such a way, the data clearly show that $r$ oscillates about a smooth, slowly varying background which {\em does not} change with $\theta$. 
This conclusion is further supported by the existence of common crossing points (cf.\,$\downarrow,\uparrow$) which occur at $\edc \simeq n \pm 1/4\,(n=1,2,3,...)$ and where $\delta \rho \simeq 0$, as prescribed by \req{hiro}.
We thus conclude that $1/\ttr$ does not change significantly in our 2DES under $B_\parallel \lesssim 1$ T.

The backscattering rate $1/\tpi$ appearing in \req{hiro} is essentially a scattering rate from the sharp disorder potential, {\em e.g.}, residual background impurities and/or interface roughness. 
Scattering off of background impurities can hardly be affected by $B_\parallel$ because of their 3D character. 
It is also unlikely that $B_\parallel$ can {\em reduce} interface roughness scattering.
Finally, a significant decrease in $1/\tpi$ should lead to a noticeable {\em decrease} in $1/\ttr$ which is not observed.
We thus conclude that the decay originates from an increase of the quantum scattering rate $1/\tq$ under applied $B_\parallel$.

%%%%%%%%%%%%%%%%%%%%%%%%%%%%%%%%%%%%%%%%%%%%%%%%%
%fig 2
%%%%%%%%%%%%%%%%%%%%%%%%%%%%%%%%%%%%%%%%%%%%%%%%%

To examine the effect of $B_\parallel$ on the quantum scattering rate $1/\tq$, we first extract the oscillatory part $\delta r/\rho_0$ by subtracting the $B_\parallel$-independent background from the data in \rfig{fig2}\,(a).
The result is presented in \rfig{fig2}\,(b) as a function of $\edc$ showing the expected period and phase with the maxima occurring near integer $\edc$ for all $\theta$.
In addition, \rfig{fig2}\,(b) allows easy extraction of the oscillation amplitude $A$ and shows, again, that the rate at which the oscillations disappear strongly depends on $\edc$, with the higher-order HIROs persisting to higher $\theta$.
%%%%%%%%%%%%%%%%%%%%%%%%%%%%%%%%%%%%%%%%%%%%%%%%%
%fig 2
\begin{figure}[t]
%\resizebox{0.5\textwidth}{!}{
\includegraphics{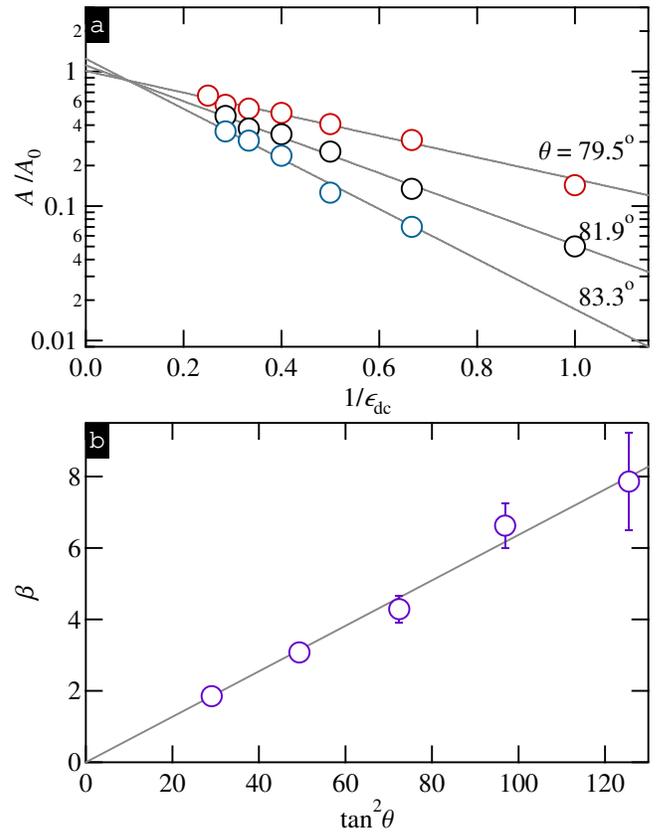}
%}
%\vspace{-0.1 in}
\caption{[Color online] (a) Normalized HIRO amplitude $A/A_0$ vs $1/\edc$ for $\theta =  79.5^\circ, 81.9^\circ$, and $83.3^\circ$ (cirles) and fits to $a\cdot\exp(-\beta/\edc)$ (lines).
(b) Extracted values of $\beta$ vs $\tan^2\theta$ (circles) and a fit, $\beta\simeq 0.064\cdot\tan^2\theta$ (line).
}
%\vspace{-0.15 in}
\label{fig3}
\end{figure}
%%%%%%%%%%%%%%%%%%%%%%%%%%%%%%%%%%%%%%%%%%%%%%%%%

We proceed with the analysis by adopting an empirical relation, $1/\tq=1/\tq^0+\Delta$, where $1/\tq^0$ is the scattering rate at $B_\parallel=0$ and $\Delta$ is the $B_\parallel$-induced correction.
Then it follows that to account for a faster decay at higher $B_\perp$, $\Delta$ should increase faster than $B_\parallel$.
Indeed, if $\Delta \propto B_\parallel$, then the argument of the Dingle factor would acquire a correction $-\pi\Delta/\oc   \propto -B_\parallel/B_\perp = -\tan \theta$.
As a result, oscillations of all orders would decay with $\theta$ at the same rate, in contradiction with our findings.
If, however, $\Delta \propto B_\parallel ^2$, then $-\pi\Delta/\oc\propto -B_\parallel^2/B_\perp \propto -\tan^2\theta/\edc$.
This correction increases with $\theta$ (for a given $\edc$) and decreases with $\edc$ (for a given $\theta$), consistent with our experimental observations.
This result implies that the amplitude should decay with $\theta$ as $A=A_0\exp(-\alpha\tan^2\theta/\edc)$, where $A_0$ is the amplitude at $\theta=0$ and $\alpha$ is a $B_\parallel$-independent constant.

In \rfig{fig3}\,(a) we present normalized HIRO amplitude $A/A_0$ vs $1/\edc$ on a semi-log scale for three different $\theta$ (as marked).
We observe that the data are described by $A=A_0\exp(-\beta/\edc)$ reasonably well and that $\beta$ increases with $\theta$.
In \rfig{fig3}\,(b) we show that $\beta$ scales roughly linearly with $\tan^2\theta$, the dependence which follows from $\Delta \propto B_\parallel^2$.
We thus conclude that the observed decay of HIROs can be explained by a $B_\parallel$-induced correction to the quantum scattering rate which scales roughly as $B_\parallel^2$.

It is interesting to examine the possibility that the $B_\parallel$-induced suppression of MIROs observed in \oncite{yang:2006} can also be explained within the same picture.
First, we notice that the Dingle factor enters equally in the description of both MIROs [\req{miro}] and HIROs [\req{hiro}] and that both MIROs and HIROs are destroyed by similar $B_\parallel \sim 1$ T. 
Second, if $B_\parallel$ is used as a parameter (instead of $\theta$), the correction to the argument of the Dingle factor can be written as $-\pi\Delta/\oc \propto -\eac\cdot\Delta$.
This correction scales with $\eac$ (and not with $1/\eac$ if $\theta={\rm const}$) and therefore {\em higher} order oscillations should be more sensitive to $B_\parallel$. 
Indeed, direct examination of Fig.\,1 in \oncite{yang:2006} reveals that (i) the number of oscillations quickly decreases with increasing $B_\parallel$, (ii) this decrease occurs at the expense of higher orders, (iii) the lower order oscillations persist to a much higher $B_\parallel$ than the higher orders, and (iv) the onset of the oscillations increases with $B_\parallel$.
All these observations imply that $1/\tq$ increases with $B_\parallel$.
We believe that a standard Dingle plot analysis (which is not feasible on our data obtained at fixed $\theta$) would readily reveal the exact dependence of $1/\tq$ on $B_\parallel$.

Finally, we mention that \oncite{yang:2006} observed that the onset of the SdHOs increased by $\simeq 50\,\%$ under applied $B_\parallel \simeq 1$ T, which indicates the same increase in the quantum scattering rate $1/\tq'$ which enters the SdHO amplitude.
Such a modest increase of quantum scattering rate, indeed, is not sufficient to explain complete quenching of MIROs.
However, $1/\tq'$ is often considerably larger than $1/\tq$ entering \reqs{miro}{hiro}, which is also 
the case for the 2DES used in \oncite{yang:2006} since the MIRO onset is much smaller than the SdHO onset.
The overestimated $1/\tq'$ is usually attributed to the fact that the amplitude of the SdHOs ($\propto \lambda^1$) is very sensitive to macroscopic density inhomogeneities which, however, do not significantly affect the amplitude of ``induced'' oscillations ($\propto\lambda^2$).
Since inhomogeneities can hardly be affected by $B_\parallel$, a modest increase in $1/\tq'$ observed in \oncite{yang:2006} likely signals a {\em much larger} increase in $1/\tq$ which, in turn, is responsible for the decay of MIROs observed in \oncite{yang:2006}. 

%\section{Summary}
In summary, we have studied the effect of an in-plane magnetic field on Hall field-induced resistance oscillations in a high mobility 2DES.
We have found that while the oscillation frequency remains unchanged, the amplitude quickly decays as the magnetic field is tilted away from the normal to the 2DES. 
The decay is very sensitive to the oscillation order and cannot be readily explained by existing theories of nonlinear transport.
Our analysis shows that the decay can be understood in terms of the $B_\parallel$-induced increase of a single particle scattering rate.
However, the exact mechanism of such an increase remains a subject of future theoretical and experimental studies.

We thank I. Dmitriev, A. Kamenev, M. Khodas, I. Kukushkin, and B. I. Shklovskii for discussions and S. Hannas, G. Jones, J. Krzystek, T. Murphy, E. Palm, J. Park, and, especially, D. Smirnov for technical assistance.
A portion of this work was performed at the National High Magnetic Field Laboratory, which is supported by NFS Cooperative Agreement No. DMR-0654118, by the State of Florida, and the DOE.
The work at Minnesota was supported by the NSF Grant No. DMR-0548014 and by the DOE Grant DE-SC0002567. 
A.T.H. acknowledges support by the Thesis Research Grant and the Doctoral Dissertation Fellowship of the University of Minnesota.

%\bibliographystyle{apsrev}
%\bibliography{../../bibliography_final_2}

\end{document}